\documentclass[nofootinbib,aps]{revtex4}
\usepackage{amssymb,amsmath,graphicx,color,microtype}
\usepackage{hyperref}
\usepackage{mathrsfs}
\usepackage{ulem}
\hypersetup{
    colorlinks=true,
    linkcolor=blue,
    filecolor=blue,
    urlcolor=blue,
    citecolor=blue
}
\usepackage[paperwidth=210mm,paperheight=297mm,centering,hmargin=2cm,vmargin=2.5cm]{geometry}
\def\be{\begin{equation}}
\def\ee{\end{equation}}
\def\ba{\begin{eqnarray}}
\def\ea{\end{eqnarray}}

\def\nn{\nonumber}
\def\lf{\left}
\def\rt{\right}

\begin{document}
\title{Derivative coupling of inflaton to $R^{(3)}$}

\author{Yan-Li He$^{1}$\footnote{heyanli16@mails.ucas.ac.cn}}
\author{Yun-Song Piao$^{1,2}$\footnote{yspiao@ucas.ac.cn}}

\affiliation{$^1$ School of Physics, University of Chinese Academy of
Sciences, Beijing 100049, China}

\affiliation{$^2$ Institute of Theoretical Physics, \\
Chinese Academy of Sciences, P.O. Box 2735, Beijing 100190, China}

\begin{abstract}

We study the inflation scenario with the non-minimally
derivative coupling $XR^{(3)}$, where $X=\nabla_\mu\phi
\nabla^\mu\phi$, $\phi$ is the inflaton and $R^{(3)}$ is the
3-dimensional intrinsic Ricci scalar on the spacelike
hypersurface,  and analytically calculate the corrections
of $XR^{(3)}$ on the power spectra of primordial perturbations. It
is found that for the $\phi^2$ inflation model, the corresponding
predictions can be driven to the best-fit region of the $n_s$-$r$
diagram.

\end{abstract}

\maketitle

\section{Introduction}

Inflation is the most popular candidate in solving the problems of
the hot Big-Bang Theory, including the flatness, the entropy and
the horizon problems as well as the monopole problem
\cite{Guth:1980zm,Linde:1981mu,Albrecht:1982wi,Starobinsky:1980te}.
Moreover, it is responsible for generating nearly scale-invariant
primordial perturbations,
 see e.g.\cite{Malik:2008im,Lyth:1998xn} for reviews.
In certain sense,  the inflation has become a standard
scenario of the early universe.

In the simplest standard slow-roll inflation case, inflaton is
just a canonical field minimally coupling to Ricci scalar
 $R$. However,  it also can be
extended to more complicated models with  the non-minimal
coupling or derivative coupling terms,  see also the
cosmological attractor models \cite{Kallosh:2013hoa,
Ferrara:2013rsa, Galante:2014ifa}.
Ref.\cite{Amendola:1993uh} discussed the coupling terms including
$XR$, $\phi^{\mu}R_{\mu\nu}\phi^{\nu}$ and $\phi\Box\phi R$, and
studied the effects of $f(\phi)R$ and $XR$ on the inflation.
Specifically, the non-minimal
coupling of the Higgs field to $R$
\cite{Bezrukov:2007ep,Bezrukov:2008ej,DeSimone:2008ei,Bezrukov:2009db,Bezrukov:2013fka,Markkanen:2017tun},
as well as the derivative coupling of the Higgs field to the
Einstein tensor $G^{\mu\nu}\phi_{\mu}\phi_{\nu}$
\cite{Germani:2010gm,Germani:2010ux}, could be used to realize
Higgs inflation. The derivative coupling
$G^{\mu\nu}\phi_{\mu}\phi_{\nu}$ also has been used in curvaton
model \cite{Feng:2013pba, Feng:2014tka}. See also, e.g.,
\cite{Huang:2015yva,Yang:2015pga,Capozziello:1999xt,Granda:2011zk,Ferreira:2018nav,Tenkanen:2017jih,Kaneta:2017lnj,Saichaemchan:2017psl,Gumjudpai:2016ioy,Artymowski:2016ikw,Broy:2016rfg,Sheikhahmadi:2016wyz,
McDonald:2015cqe,Ema:2015oaa,Arapoglu:2015xua,delCampo:2015wma,Chiba:2014sva,Goodarzi:2014fna,Chen:2014zoa,Darabi:2013caa,Sadjadi:2013psa,White:2013ufa,Kim:2013st,Artymowski:2012is,White:2012ya}
for other applications of non-minimal derivative coupling in
cosmology.


Inspired by the significant role played by the $\delta
g^{00}R^{(3)}$ operator in curing the instabilities of scalar
perturbations in nonsingular cosmology \cite{Cai:2016thi,
Creminelli:2016zwa, Cai:2017tku, Cai:2017dyi, Kolevatov:2017voe},
we propose in this paper  a new non-minimally derivative
coupled scenario in which the kinetic term of the inflaton, i.e.,
$X$, couples directly to the geometric variable $R^{(3)}$
(3-dimensional Ricci scalar). This coupling does not affect
background evolutions and only modify the spatial derivative terms
of scalar and tensor perturbations. Such a coupling model actually
belongs to a special subclass of beyond Horndeski theory \cite{
Gleyzes:2014qga, Gleyzes:2014dya, Langlois:2015cwa} (with the
absence of ${}^{H}L_5$ and ${}^{BH}L_5$),  see also
Appendix~\ref{codisformal}, so there is not the Ostrogradski
instability.

 We will calculate the effect of the derivative coupling
$XR^{(3)}$ on the spectra of primordial perturbations. Since in
unitary gauge, $X=g^{00}\dot\phi_0^2$ and $g^{00}=-1+\delta
g^{00}$, our model is also equivalent to adding operators \ba
\label{add} L_{add-oper}\sim M_p^2\frac{\tilde m_4^2(t)}{2}\delta
g^{00}R^{(3)}-M_p^2\frac{m_R^2(t)}{2}R^{(3)} \ea to standard
canonical slow-roll inflation action,
where, $\tilde
m_4^2(t)=m_R^2(t)$. We will work in the frame in which
the graviton behaves like in the standard one, i.e. the propagating
speed of graviton equals to the speed of light. By performing a
disformal transformation \ba \label{disformal} \tilde
g_{\mu\nu}=C(t)g_{\mu\nu}+D(t)n_{\mu}n_{\nu}, \ea we will get
rid of the second term in (\ref{add}). Note that the spectra of
both scalar and tensor perturbations are disformally
invariant\cite{Creminelli:2014wna, Tsujikawa:2014uza,
Watanabe:2015uqa, Domenech:2015hka}. Additionally, the
corresponding covariant Lagrangian also preserves the structure of
the beyond Horndeski theory, see also Appendix~\ref{codisformal}.

We obtain the power spectrum of scalar perturbation, as
well as the tensor perturbation, and study the impact of
$XR^{(3)}$ on the $n_s$-$r$ diagrams  of a few inflation
models. Especially, it is found that the appearance of the
$XR^{(3)}$ term with the negative coupling can drive $\phi^2$
inflation to the best-fit region of the $n_s$-$r$ diagram.

\section{Derivative coupling of inflaton to $R^{(3)}$}

\subsection{The covariant theory} \label{the covariant}

As introduced, we will study the inflation scenario with
the action \ba S=\int
dx^4\sqrt{-g}\frac{M_p^2}{2}\lf[R-X-2V(\phi)+\frac{2}{M_p^2}L_{XR^{(3)}}\rt],
\label{cova1} \ea where
$X=\phi_{\mu}\phi^{\mu},~\phi_{\mu}=\nabla_{\mu}\phi,~\phi^{\mu}=\nabla^{\mu}\phi$,
$\phi$ is the dimensionless inflaton and
$L_{XR^{(3)}}$ is the covariant expression of $XR^{(3)}$.

We will derive the covariant expression $L_{XR^{(3)}}$. Adopting the Gauss-Codazzi relation, it is
straightforward to find \ba R^{(3)}&=&
R-{\phi_{\mu\nu}\phi^{\mu\nu}-(\Box \phi)^2\over X} +{2 \over
X^2}(\phi^\mu\phi_{\mu\nu}\phi^{\nu\sigma}\phi_\sigma-\phi^\mu
\phi_{\mu\nu}\phi^\nu \Box \phi)
-{2R_{\mu\nu}\phi^\mu\phi^\nu\over X}\,, \ea where the last term
can be recast as \ba \phi^{\mu}R_{\mu\nu}\phi^{\nu}&=& {
\phi^{\mu}\nabla_{\nu}\nabla_{\mu}\phi^{\nu}
}-\phi^{\mu}\nabla_{\mu}\nabla_{\nu}\phi^{\nu}. \label{der3} \ea
By integration by parts, we have \cite{Cai:2017dyi} \ba
\label{LXR} L_{XR^{(3)}}&=&\frac{f_{1}}{2}XR^{(3)},\nn\\
&=&\frac{f_1}{2}XR+\frac{f_1}{2}(\phi_{\mu\nu}\phi^{\mu\nu}-\Box\phi^2)+\frac{f_1}{X}(\phi^{\mu}\phi_{\mu\nu}\phi^{\nu\sigma}\phi_{\sigma}-\Box\phi\phi^{\mu}\phi_{\mu\nu}\phi^{\nu})\nn\\
&&-\frac{f_{1\phi\phi}}{2}X^2-\frac{3}{2}f_{1\phi}X\Box\phi\,,
\ea
where the
subscript $\phi$ denotes the derivative with respect to $\phi$.

The covariant $L_{XR^{(3)}}$, which contains quadratic order of the
second order derivative of $\phi$ and the lowest order derivative
of $\phi$ coupling to gravity (i.e.$XR$), actually belongs to a
subclass of the beyond Horndeski theory \cite{Gleyzes:2014dya}
(see Appendix \ref{codisformal} for details). A combination of
$^{H}L_4$ and $^{BH}L_4$ is degenerate, which leads to
the absence of Ostrogradski instability \cite{Langlois:2015cwa}.
 It also should be pointed out that $L_{XR^{(3)}}$ does not
affect the background evolution.

\subsection{The EFT of cosmological perturbations}\label{disformal}
 In the EFT approach of inflation \cite{Cheung:2007st}, the
action (\ref{cova1}) actually corresponds to \ba\label{Seft}
S=\int
dx^4\sqrt{-g}\lf[\frac{M_p^2}{2}R-c(t)g^{00}-\Lambda(t)+M_p^2\frac{f(t)}{2}\delta
g^{00}R^{(3)}-M_p^2\frac{f(t)}{2}R^{(3)}\rt], \ea where
$\Lambda(t)=V(\phi_0(t))M_p^2$,~~$c(t)=\frac{1}{2}\dot\phi_0^2(t)M_p^2$,
$f(t)=f_1(\phi(t))\frac{f_2(t)}{M_p^2}$, $f_2(t)={\dot
\phi_0}^2(t)$ and $f(t)$ is dimensionless. Action (\ref{Seft}) is
equivalent to GR plus the canonical field and the set of operators
in (\ref{add}) when $\tilde
m_4^2=f(t),~m_R^2=f(t)$. As noted in Ref.\cite{Bordin:2017hal},
the operator $R^{(3)}$ modifies the coupling
$\langle\gamma\gamma\rangle$, i.e., tensor fluctuations.

The Fridmann equations are given by
\ba \label{back1} H^2&=&\frac{1}{3M_p^2}\lf(\Lambda(t)+c(t)\rt),\\
\dot H+H^2&=&\frac{1}{3M_p^2}\lf(\Lambda(t)-2c(t)\rt),\ea
where a dot represents the time derivative with respect to $t$.

We can write the action $\int dx^4\sqrt{-g}\left(\frac{M_p^2}{2}R+L_{XR^{(3)}}\right)$ as \ba \int dx^4\sqrt{-g}\frac{M_p^2}{2}\left[(1-f(t))R^{(3)}+K_{\mu\nu}K^{\mu\nu}-K^2+f(t)\delta
g^{00}R^{(3)}\right]\,, \label{action01} \ea where we have used
the Gauss-Codazzi relation \ba\label{RR3}
R=R^{(3)}-K^2+K_{\mu\nu}K^{\mu\nu}-2\nabla_{\mu}\lf(A^{\mu}-Kn^{\mu}\rt),
\ea with $A^{\mu}=n^{\rho}\nabla_{\rho}n^{\mu}$ being the
acceleration vector and $n_{\mu}=-\partial_{\mu}\phi/\sqrt{-X}$ being
the normal vector perpendicular to the hypersurfaces.

 It is convenient to calculate the perturbations in the
frame in which the graviton behaves like in GR apparently,
or see e.g.\cite{Cai:2015yza, Cai:2016ldn}. For this purpose, we consider a
field redefinition of $g_{\mu\nu}$ consisting of a conformal
 rescaling and a lightcone structure-disformal term on the
four-dimensional spacetime manifold \cite{Bordin:2017hal} \ba
\label{metric}\tilde
g_{\mu\nu}=C(t)g_{\mu\nu}+D(t)n_{\mu}n_{\nu}\,, \ea \ba \tilde
g^{\mu\nu}=\frac{1}{C}g^{\mu\nu}+\frac{D}{C(D-C)}n^{\mu}n^{\nu}.
\ea  Such a redefinition can be used to apparently get rid of the effect of
 the term $f(t)R^{(3)}$ in action (\ref{Seft}) on the tensor perturbation.
Meanwhile, it redefines the scale factor and the cosmic time of
the background FRW spacetime, but does not affect the
power spectra of both scalar and tensor perturbations.

Since the metric only determines the coefficient of the normal
vector, and the foliation of spacetime remains unchanged, $\tilde n_{\mu}$ should be parallel to $n_{\mu}$. We
define $\tilde n_{\mu}=B(t)n_{\mu},~\tilde
n^{\mu}=\lf(B(t)\rt)^{-1}n^{\mu}$. After some simple calculations,
we obtain $B=\sqrt{C-D}$. Furthermore, in the unitary gauge, we
recall that $n_{\mu}=-N\delta_{\mu}^0$, which indicates $\tilde
N^2=(C-D)N^2$. According to the definition of the induced metric
$\tilde h_{\mu\nu}$ with respect to $\tilde g_{\mu\nu}$, i.e., \ba
\label{h}\tilde h_{\mu\nu}&=&\tilde g_{\mu\nu}+\tilde
n_{\mu}\tilde n_{\nu}, \ea it is easy to find that $\tilde
h_{\mu\nu}=Ch_{\mu\nu}$,  which suggests \be \label{R3}
R^{(3)}=C\tilde R^{(3)} \ee and $\tilde N^i=N^i$. The relation
between the determinant of two induced metrics $\tilde h=C^3h$ combined with $\tilde N=\sqrt{C-D}N$  suggest that $\sqrt{-\tilde g}=C^{3/2}\sqrt{C-D}\sqrt{-g}$.

The extrinsic curvature $K_{\mu\nu}$ obeys \be
\label{K}K_{\mu\nu} =\frac{\sqrt{C-D}}{C}\lf(\tilde
K_{\mu\nu}+\frac{1}{2}\tilde h_{\mu\nu}\sigma\rt), \ee where
$\sigma =-{\mathscr L}_{\tilde {\textbf{n}}}\ln C$, and $\mathscr L$ is the Lie derivative with respect to $\tilde n^{\mu}$.


We also need to perform the rescaling of the time
coordinate $t$  \ba \label{transformt}
dt=\frac{1}{\sqrt{C-D}}d\tilde t, \ea  where $C(t)$
and $D(t)$ depend only on time, so that the metric of the
background spacetime after the transformation remains
flat FLRW, which implys that $\tilde N=N$. By some manipulations, we can obtain $\sigma=\frac{\alpha}{\tilde N}$, where \ba
\label{a}\alpha=-\frac{d C}{d\tilde t}\frac{1}{C}.
\ea
Neither the covariant volume element that is diffeomorphism invariant nor $\tilde R^{(3)}$ and the extrinsic curvature associated with the foliation of the spacetime are affected by the time rescaling.


With Eqs.(\ref{R3}), (\ref{K}) and (\ref{transformt}), we
can rewrite (\ref{action01}) as
\begin{eqnarray}
&& \int d\tilde t dx^3\sqrt{-\tilde g}\frac{1}{\sqrt{C-D}}\frac{M_p^2}{2}\bigg[C^{-\frac{1}{2}}\lf(1-f(t(\tilde t))\rt)\tilde R^{(3)}+C^{-\frac{3}{2}}\lf(C-D\rt)\lf(\tilde K_{\mu\nu}\tilde K^{\mu\nu}-\tilde K^2\rt)\nn\\
&&+C^{-\frac{1}{2}}f(t(\tilde t))\delta{\tilde g}^{00}\tilde {R}^{(3)}-C^{-\frac{3}{2}}(C-D)\lf(2\sigma\tilde K+\frac{3}{2}\sigma^2\rt)\bigg].
\end{eqnarray}
Requiring the coefficients of $\tilde R^{(3)}$ and
$\tilde K_{\mu\nu}\tilde K^{\mu\nu}-\tilde K^2$ being unity sets the
values of $C$ and $D$. As a result, (\ref{metric})  can be written
as \ba \label{transformation} \tilde
g_{\mu\nu}=C g_{\mu\nu}+Dn_{\mu}n_{\nu} \,, \ea where
\ba C(t(\tilde t))=\sqrt{1-f(t(\tilde t))},~~D(t(\tilde
t))=f(t(\tilde t))\sqrt{1-f(t(\tilde t))}\,. \ea
It is apparent that $D=C(1-C^2)$. With
transformation (\ref{transformation}), the coefficient of Einstein-Hilbert term
is recast in the standard form.

Additionally,  $\tilde h_{ij}=Ch_{ij}$ and
Eq.(\ref{transformt}) will give the relations between $H$ and
$\tilde H$ as following, \ba
\label{H} H&=&\lf(1-f\rt)^{\frac{3}{4}}\lf(\tilde H+\frac{1}{2}\alpha\rt),\\
\label{dotH} \frac{dH}{dt}&=&(1-f)^{\frac{3}{2}}\lf(\frac{d\tilde H}{d\tilde t}+\frac{1}{2}\frac{d\alpha}{d\tilde t}-\frac{3}{2}\alpha\tilde H-\frac{3}{4}\alpha^2\rt).
\ea

Thus, up to the second order of the EFT operators, Eq.(\ref{Seft})
 can be written as \ba
\label{actioneft2}
 &\,&{\tilde S}=\int d\tilde t dx^3\sqrt{-\tilde g} \frac{M_p^2}{2}\Big[\tilde R-2( \dot{\tilde H}+3\tilde H^2)+2\dot {\tilde H}\tilde g^{00}\nn\\ &\,&\qquad\qquad\qquad\qquad\qquad
 +\alpha_f\delta\tilde g^{00}\tilde R^{(3)}
+\frac{1}{4}(3\alpha\tilde H-\dot\alpha){\delta {\tilde
g^{00}}}^2+\alpha\delta {\tilde g^{00}}\delta {\tilde K}\Big], \ea
where \ba \label{af} \alpha_f=\frac{f}{1-f}\ea is dimensionless.
Using (\ref{a}), $\alpha$ is related to $f$ by \ba
\alpha=\frac{1}{2}\frac{1}{1-f}\frac{df}{d\tilde t}\, \label{a2}.
\ea Here, and throughout the rest of the paper, dot represents
time derivative with respect to $\tilde t$. In Appendix
\ref{codisformal}, we will demonstrate that action
(\ref{actioneft2}) can also be obtained in  covariant
language.



In Eq.(\ref{actioneft2}), the first three terms are expectantly
dependent on the background evolution, while the remainder starts
from quadratic order in the perturbations.
In the original frame, the
graviton has a nontrivial sound speed $c_T^2=1-f(t)$; In  the
new frame,  the graviton behaviors as in the GR, and the
main contributor to the sound speed squared of scalar perturbation
$c_s^2$ is $\delta \tilde g^{00}\tilde R^{(3)}$. For a constant
$f$, $\alpha=0$, action (\ref{actioneft2}) is equivalent to the standard slow-roll
inflation case plus the operator $\delta {\tilde g}^{00}\tilde
R^{(3)}$.

\section{The power spectrum of primordial perturbation}

\subsection{Background equations}

In this section, let's derive the background equations in the new
transformed frame. The background spacetime is flat FLRW. From the covariant action (\ref{covs02}), we obtain the modified Friedmann equations \ba
\label{back01} 3\tilde H^2&=&\frac{1}{2}\dot\phi^2+U(\phi)-\frac{3}{4}\alpha^2-3\alpha\tilde H,\\
\label{back02} \dot{\tilde
H}&=&-\frac{1}{2}\dot\phi^2+\frac{3}{4}\alpha^2-\frac{1}{2}\dot\alpha+\frac{3}{2}\alpha\tilde
H \ea and the equation of motion of $\phi$ \ba \label{phimotion}
\dot\phi \left(\ddot\phi+3\tilde
H\dot\phi+U_{\phi}\right)-\frac{3}{2}\alpha\lf(\dot\alpha+3\alpha
\tilde H\rt)-3\alpha \tilde H^2\lf(3-\tilde\epsilon\rt)=0, \ea
where $U$ is the effective potential in the new frame, \ba
U(\phi)=V(1+\alpha_f)^{\frac{3}{2}} \label{U}. \ea

Besides the standard slow-roll conditions $\dot\phi^2\ll U$ and
$\vert\ddot\phi\vert\ll 3\tilde H\vert\dot\phi\vert$, we still
need additional slow-roll conditions $\vert\frac{\alpha}{\tilde
H}\vert\ll 1$ and $\vert\dot\alpha\vert\ll \vert3\alpha\tilde
H\vert$.
Hence, up to first order in slow-roll parameters, the background
equations are approximately rewritten as \ba
\label{slow1} 3\tilde H^2&\simeq&U,\\
\label{slow2} \dot{\tilde H}&\simeq&-\frac{1}{2}\dot\phi^2+\frac32\alpha\tilde H,\\
\label{slow3} 3\tilde
H\dot\phi&+&V_{\phi}(1+\alpha_f)^{\frac32}\simeq0. \ea
 The number of e-folds is computed
as follow \ba
N(\phi)&\simeq&\int^{\phi}_{\phi_{end}}\frac{V}{V_{\phi}}d\phi,\label{N}
\ea

\subsection{Primordial perturbations}

The quadratic action of scalar perturbation for (\ref{actioneft2})
is
$$S_{\zeta}^{(2)}=\int d\tilde t dx^3\tilde a^3\left[c_1\dot\zeta^2-(\frac{\dot c_3}{\tilde a}-M_p^2)\frac{(\partial\zeta)^2}{\tilde a^2}\right],$$
where
\begin{eqnarray}
c_1&=&\frac{M_p^2}{(2+\tilde\eta)^2}\lf(4\tilde\epsilon+6\tilde\eta+2\tilde\epsilon\tilde\eta+3\tilde
\eta^2-2\tilde\eta\tilde\eta_1\rt),\\
c_3&=&\frac{2\tilde aM_p^2}{(2+\tilde\eta)\tilde
H}\lf(1+2\alpha_f\rt)\,.
\end{eqnarray}
The slow-variation parameters $\tilde\epsilon$,~~$\tilde\eta$ and
$\tilde\eta_1$ are given by \ba
\label{slowroll01} &&\tilde\epsilon=-\frac{\dot{\tilde H}}{\tilde H^2},~~\tilde\epsilon_1=\frac{\dot{\tilde\epsilon}}{\tilde H\tilde\epsilon},~~\tilde\epsilon_n=\frac{d\ln\tilde\epsilon_{n-1}}{\tilde Hd\tilde t}(n>1),\\
\label{slowroll02} &&\tilde\eta=\frac{\alpha}{\tilde
H},~~\tilde\eta_1=\frac{\dot{\tilde\eta}}{\tilde
H\tilde\eta},~~\tilde\eta_n=\frac{d\ln\tilde\eta_{n-1}}{\tilde Hd\tilde
t}(n>1), \ea see a hierarchy of Hubble flow parameters in
\cite{Schwarz:2001vv, Leach:2002ar}. From Eq.(\ref{a}), \be
\tilde\eta=-\frac{\dot C}{\tilde
HC}=-\alpha_C\frac{\dot\phi}{\tilde H}. \ee where,
\be\alpha_C=\frac{C_{\phi}}{C}\label{alphac}.\ee During slow-roll
inflation, the slow-roll parameters $\epsilon_n\ll 1$, as well as
$\eta_n$.

The sound speed squared reads
$$c_s^2=\lf(\frac{\dot c_3}{\tilde a}-M_p^2\rt)/c_1,$$
which can be rewritten as
\begin{eqnarray}
\label{cs2}c_s^2&=&1+2\alpha_f+\frac{4\tilde\eta(2+\tilde\eta)+2\alpha_f(2+\tilde\eta)^2}
{4\tilde\epsilon+6\tilde\eta+2\tilde\epsilon\tilde\eta+3\tilde\eta^2-2\tilde\eta\tilde\eta_1}.
\end{eqnarray}
Here, $\delta {\tilde g}^{00}\tilde R^{(3)}$ modifies $c_3$ and
also $c_s^2$. Apparently, $c_s^2>0$ and $c_1>0$ are required to avoid the
small-scale Laplacian instability and ghost instability. The key factor which causes the sound speed of scalar
perturbation to deviate from unity is $\alpha_f$, namely the
coefficient of the operator $\delta \tilde g^{00}\tilde
R^{(3)}$.

The equation of motion for perturbation is \ba \label{pmotion}
u''+\lf(c_s^2k^2-\frac{z_s''}{z_s}\rt)u=0, \ea where
$u=z_s\zeta$,~~$z_s=\sqrt{2\tilde a^2c_1}$, the superscript $'$ is
the derivative with respect to the conformal time $\tilde\tau$,
and $\tilde\tau=\int d\tilde t/\tilde a$.

In the following, we will analytically estimate the power spectrum
of the scalar perturbation. In analogy with
Ref.\cite{DeFelice:2014bma}, we define the following slow roll
parameters \ba \label{slowroll2} \epsilon_s=\frac{\dot c_s}{\tilde
Hc_s},~~\delta=\frac{\dot c_1}{\tilde Hc_1}, \ea which are much
less than unity. If $\epsilon_{s}$ does not satisfy this condition,
$c_s^2$ may have moderately sharp features, which may disrupt
slow-roll.

 We define a new evolution parameter $y$ by
$
y= c_s\tau,
$
 whose time derivative is
$
\frac{dy}{d\tau}=c_s(1- \epsilon_s )
$.
After some calculations, Eq.(\ref{pmotion}) can be recast as
\be
 (1-2\tilde\epsilon-2\epsilon_s) u_{yy} - {\epsilon_s}\frac{1}{y}u_y +\lf(k^2 {(1-2\tilde\epsilon)}-\frac{1}{y^2}(2-\tilde\epsilon+\frac32\delta)\rt)u=0,\label{ppmotion}
\ee
where and in the following, we  ignore the slow-roll corrections of the order of $\tilde\epsilon^2$ or non-linear order corrections.
The solution of Eq.(\ref{ppmotion}) is
\begin{align}
u=y^{\frac{1+\epsilon_s}{2}}
\Big[C_1H^{(1)}_{\nu_s}\lf(-(1+\epsilon_s)ky\rt)
+C_2H^{(2)}_{\nu_s}\lf(-(1+\epsilon_s)ky\rt)\Big],\label{solution1}
\end{align}
in which $H^{(1)}_{\nu_s}(y)$ and $H^{(2)}_{\nu_s}(y)$ are the first and second kinds Hankel functions of $\nu_s$-th order, respectively, $C_1$ and $C_2$ are two constants, and
\be
\nu_s\simeq\frac{3}{2}+\tilde\epsilon+\frac{1}{2}\delta+\frac{3}{2}\epsilon_s.
\ee

The initial state of perturbation mode is
$u\simeq \frac{1}{\sqrt{2c_sk(1+\epsilon_s)}}{e^{-i(1+\epsilon_s)ky}} $ for $-ky\gg1$
in the Bunch-Davies vacuum. In addition, when $-ky\rightarrow +\infty$,
\begin{align}
H^{(1)}_{\nu_s}\lf[-(1+\epsilon_s)ky\rt]&\rightarrow -\sqrt{\frac{2}{-\pi(1+\epsilon_s) ky}}e^{i(-(1+\epsilon_s)ky+(3-2\nu_s)/4\pi)},\label{H1}\\
H^{(2)}_{\nu_s}\lf[-(1+\epsilon_s)ky\rt]&\rightarrow -\sqrt{\frac{2}{-\pi(1+\epsilon_s) ky}}e^{i((1+\epsilon_s)ky-(3-2\nu_s)/4\pi)},
\end{align}
therefore we have $C_2 = 0$. With this condition, Eq.(\ref{solution1}) is recast as
\begin{align}
u& =C_1y^{\frac{1+\epsilon_s}{2}}H^{(1)}_{\nu_s}\lf(-(1+\epsilon_s)ky\rt), \label{solution}
\end{align}
Using the property (\ref{H1}),
the solution in the asymptotic past reads
\be
u\simeq -C_1y^{\frac{1+\epsilon_s}{2}} \sqrt{\frac{2}{-\pi(1+\epsilon_s)ky}}e^{i\lf(-(1+\epsilon_s)ky-\frac{\pi}{2}\nu_s+\frac{3}{4}\pi\rt)}.
\ee
$C_1$ is determined by the Wronskian normalization $i=\frac{\mathrm{d}y}{\mathrm{d}\tau}\left(u\frac{\partial u^{\ast}}{\partial y}-u^{\ast}\frac{\partial u}{\partial y}\right)$,
which means
\be
C_1=\frac i2\sqrt{\frac{\pi}{ c_{s\ast}}}(1+\frac{1}{2}\epsilon_s)y_{\ast}^{-\frac{\epsilon_s}{2}},\label{c1}
\ee
where, $c_{s\ast}$ is the value of $c_s$ at horizon crossing $c_sk=aH$ (i.e., at $y=y_{\ast}$), and
\be
y_{\ast}=c_{s\ast}\tau_{\ast},\quad c_s=c_{s\ast}\left(\frac{y}{y_{\ast}}\right)^{- {\tilde\epsilon_s}/{\mu}},\quad\mu=1-\tilde\epsilon-\epsilon_s.
\ee
Therefore, from Eqs.(\ref{solution}) and (\ref{c1}), we obtain
\ba
u =\frac 12\sqrt{\frac{-\pi y}{ c_{s }}}(1+\frac{1}{2}\epsilon_s) H^{(1)}_{\nu_s}\lf(-(1+\epsilon_s)ky\rt).
\ea

On super horizon scales, i.e. for long wavelength perturbations ($-ky\ll1)$,
\be
H_{\nu_s}^{(1)}(-(1+\epsilon_s)ky)\simeq e^{-i\frac{\pi}{2}}\left(-\frac{2}{(1+\epsilon_s)ky}\right)^{\nu_s}\frac{\Gamma(\nu_s)}{\pi},
\ee
where, $\Gamma(x)$ denotes the Gamma function. Up to  leading-order corrections, the power spectrum of scalar
perturbation is
\begin{eqnarray}
P_s(k)&=&\frac{k^3}{2\pi^2}\lf|\frac{u(k)}{z_s}\rt|^2,\\
&=&\frac{\tilde H^2}{2\pi^3}\frac{ \left(\Gamma(\nu_s)\right)^2
}{c_1c_s^3}\left(-\frac{ky}{2}\right)^{3-2\nu_s}.
\end{eqnarray}

The spectral index is defined through the scale dependence of the
power spectrum
\begin{align}
n_s-1=&\frac{d\ln P_s(k)}{d\ln k},\\
=&-2\tilde\epsilon-\delta-3\epsilon_s.
\end{align}
The tensor-to-scalar ratio is approximately \ba
r=\frac{P_t(k)}{P_s^{lead}(k)}\simeq16\tilde\epsilon c_s^3,
\ea  where $P_t(k)$ is the standard power spectrum of
tensor perturbation. The standard consistency relation between
$r$ and $n_t$ is broken due to the presence of $\alpha_f$ in the
slow-roll inflation with the $XR^{(3)}$ correction.  The tensor-to-scalar ratio is
suppressed for a negative $\alpha_f$ while it is enhanced for a
positive $\alpha_f$.

Up to linear order of  the slow-roll parameter, the
slow-roll parameters in new frame are related to counterparts in
the original frame by \ba
\tilde\epsilon&\simeq&\epsilon-\frac{3}{2}\eta\simeq\epsilon,\label{epsilon}\\
\tilde\epsilon_1&\simeq&\epsilon_1,\label{epsilon1}\\
\tilde\eta&=&\frac{2\eta}{2-\eta}\simeq\eta,\label{eta} \ea
where $\eta=\frac12\frac{1}{(1-f)H}\frac{df}{dt}$.  
We assuming that $\alpha_f\lesssim{\cal O}(\epsilon)$ (thus
$f\lesssim{\cal O}(\epsilon)$) and $\eta\lesssim{\cal
O}(\epsilon^2)$. Up to first order corrections, with
Eqs.(\ref{epsilon}) and (\ref{eta}), (\ref{cs2}) can be
written by \ba
c_s^2&\simeq&1+2\alpha_f+\frac{2}{\epsilon}(\eta+\alpha_f)\\
&\simeq&1+2f\lf(1+\frac{1}{\epsilon}\rt)+\frac{2}{\epsilon}\lf(\eta+f^2\rt)\,.
\ea
Up to the first order of slow-roll parameters, $c_1\simeq \epsilon$,
thus $\delta\sim \epsilon_1$. Using
(\ref{slowroll01}), (\ref{slowroll02}) and
(\ref{epsilon})-(\ref{eta}), we can recast (\ref{slowroll2}) as \ba \label{epsilons}
\epsilon_s&\simeq&\frac{2\eta-f\epsilon_1 }{\epsilon+2f}.
\ea
Employing the original slow roll parameters and
Eq.(\ref{epsilons}), we have \ba n_s-1 \simeq
-2\epsilon-\epsilon_1-3\frac{
2\eta-f\epsilon_1}{\epsilon+2f},\label{spectral} \ea which shows
that the spectral index of scalar perturbation contains not only
the Hubble flow parameters but also the slow-roll parameters defined
by the time derivative of $f$. For $\eta=0$, the spectral index is
modified due to $\alpha_f\simeq f\neq0$. The power spectrum of the
gravitational waves is unaffected by $f$ and its time
derivative; thus, its spectrum is still the standard result like in
GR, which is consistent with the observations.

Similarly, up to first order corrections, we have \ba
r&\simeq&16\epsilon\lf(1+\frac{2f}{\epsilon}\rt)^{\frac32}\left(1+2\frac{f\epsilon+f^2+\eta}{\epsilon+2f}\right)^{\frac{3}{2}},\\
&\simeq&16\epsilon\lf(1+\frac{2f}{\epsilon}\rt)^{\frac32}\left(1+3\frac{f\epsilon+f^2+\eta}{\epsilon+2f}\right).
\ea

The original slow roll parameters can be expressed in terms of the
effective potential and the $XR^{(3)}$ coupling function \ba
\epsilon&\simeq&\frac{V_{\phi}^2}{2V^2}\label{epsilonv},\\
\epsilon_1&\simeq&-2\lf(\frac{V_{\phi\phi}}{V}-\frac{V_{\phi}^2}{V^2}\rt),\label{epsilon1v}\\
\eta&=&\frac{1}{2}\frac{f_{\phi}}{
H(1-f)}\frac{d\phi}{dt}\simeq\alpha_C\frac{V_{\phi}}{V}\label{etav},
\ea where $\alpha_C=-\frac{1}{2}\frac{f_{\phi}}{1-f}$.

\section{Inflation models with $V\sim\phi^n$}
In this section, in order to illustrate  the impact of
$XR^{(3)}$ , we will consider the inflation models with
$V\sim\phi^n$, but with different forms of the coupling
coefficient $f_1$, including a power-law $f_1$ and a dilaton-like
$f_1$.

\subsection{Power-law coupling
coefficient}
We consider the model in which
\ba \label{power} f_1&=&f_0\phi^{-n},~~~V=V_0\phi^n
\ea
with $f_0$ and $V_0$ being constants. By imposing
(\ref{slow1}), (\ref{slow2}) and (\ref{slow3}), one gets \ba
f&=&\frac{1}{3}n^2\beta\phi^{-2},~~~f_{\phi}=-\frac{2}{3}n^2\beta\phi^{-3},
\ea where $\beta\equiv f_0V_0$. In this model, the slow-roll
parameters are given by \ba
\epsilon&=&\frac{1}{2}n^2\phi^{-2},\\
\epsilon_1&=&2n\phi^{-2},\\
\eta&=&\frac{1}{3}\beta n^3\phi^{-4}. \ea

The power spectral index is given by \be n_s-1 =
-\frac{2(n+2)}{4N+n}. \ee
Note that the tensor-to-scalar ratio depends on $\beta$, but the
spectral index does not. This is mainly because that the last term
$2\eta-\epsilon\alpha_f$ vanishes in Eq.(\ref{spectral}). The
scalar spectrum index is independent of $\beta$ and $\eta$ up to
the first order in the slow roll approximation.

\begin{figure}[htbp]
	\includegraphics[scale=2,width=0.7\textwidth]{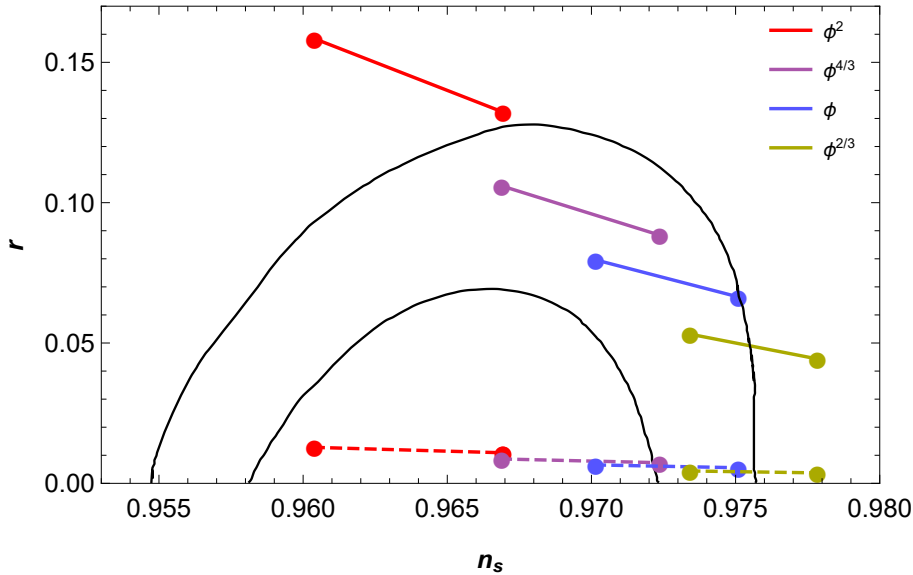}
	\caption{ The $n_s$-$r$ points predicted by the model (\ref{power}) for different
			parameters are confronted with Planck observation.
		The red, purple, blue and green lines correspond to  the $\phi^n$
			models with
			$n=2$, $n={4}/{3}$, $n=1$  and $n={2}/{3}$, respectively.
		The solid lines correspond to the $n_s$-$r$  values of the
			the models without $XR^{(3)}$. The dashed lines correspond to the
			shifted $n_s$-$r$ values in the corresponding models with the
			$XR^{(3)}$ coupling with $\beta=-0.60$.
		The dots on the left hand side and on the right hand side correspond
		to the e-folding number $N=50$ and $N=60$, respectively.}
	\label{figbeta}
\end{figure}

 Up to first order correction, we
plot the consistency relations predicted by the inflation models
with $\beta=0$ (i.e. without the $XR^{(3)}$ coupling) and
compare the results with that of $\beta=-0.60$ on the $n_s$-$r$
diagram in Fig.\ref{figbeta}. Marginalized joint $\sigma$,
$2\sigma$ contours from inside to outside for $n_s$ and $r$
 are plotted according to the Planck 2018 data. The dashed
lines are the predictions of the modified consistency relations,
while the solid lines are the standard consistency relations. The
parameter $\beta$ can shift the predicted $r$ vertically for a
fixed number of e-folding, and  $\beta<0$ leads to a
reduced tensor-to-scalar ratio.

As we can see from Fig.\ref{figbeta}, each model has a
smaller tensor-to-scalar ratio after considering $XR^{(3)}$
corrections.  Moreover, compared with the models with
$n=1,2/3$, the model with $n=2$, i.e.$\phi^2$ model, can be driven
to the best-fit region favored by the observation.

\subsection{Dilaton-like coupling coefficient}

Now, we consider the case with a dilaton-like coupling
coefficient \ba \label{exponential}
f_1&=&f_0e^{-\lambda\phi},~~~V=V_0\phi^n. \ea Similar to the
previous model, one gets \ba f=\frac{1}{3}\beta
n^2\phi^{n-2}e^{-\lambda\phi},~~~f_{\phi}=\frac{1}{3}\beta
n^2\phi^{n-3}e^{-\lambda\phi}(n-2-\lambda\phi). \ea The Hubble
flow parameter $\epsilon$ is same with the previous model, but \ba
\eta=-\frac{1}{6}\beta
n^3\phi^{n-4}e^{-\lambda\phi}(n-2-\lambda\phi). \ea The spectral
index $n_s$  can be written
as \ba
n_s-1&=& -n(n+2)\phi^{-2}+\frac{2\beta n\phi^{n-2}e^{-\lambda\phi}(n-\lambda\phi)}{1+\frac{4}{3}\beta\phi^ne^{-\lambda\phi}},\\
\nn \ea which involves model parameters $\beta$ and $\lambda$ in the
slow-roll approximation.

\begin{figure}[htbp]
	\centering
	\includegraphics[scale=2,width=0.7\textwidth]{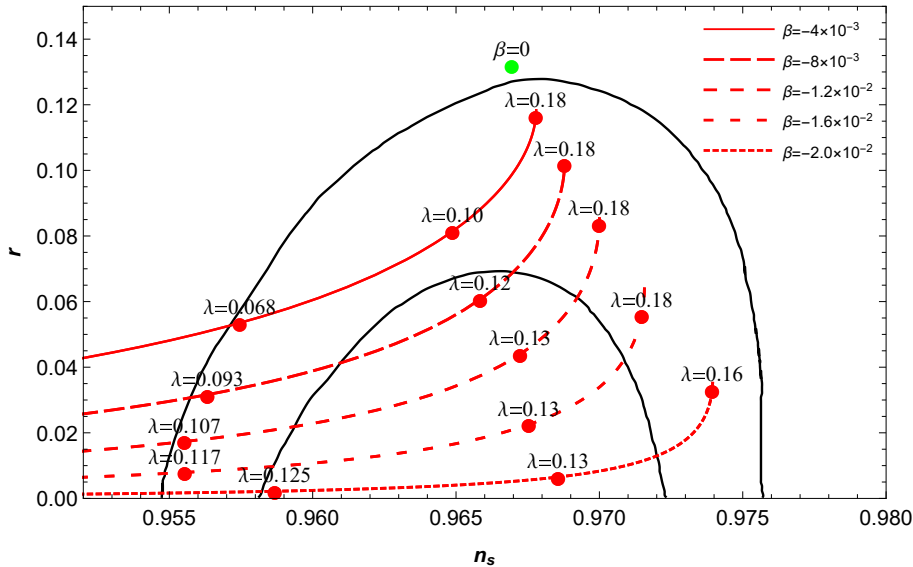}
	\caption{ The $n_s$-$r$ points predicted by the model (\ref{exponential})
			for different values of $\lambda$ and $\beta$ are
			confronted with Planck observation.
		Here we choose $N=60$.  The green dot corresponds
			to the case with $\beta=0$, i.e., $\phi^2$ model without the $XR^{(3)}$ coupling.}
	\label{figAnovanish}
\end{figure}

 We restrict ourselves to the $\phi^2$ model with $N=60$.
The prediction of $n_s$-$r$ with the different values of the
parameters $\beta$ and $\lambda$ is plotted in
Fig.\ref{figAnovanish}. The green dot corresponds to $\beta=0$,
which is the prediction of standard consistency relation without
the coupling $XR^{(3)}$. From top to bottom, the red curves correspond to
$\beta=-4\times 10^{-3}, -8\times10^{-3}, -1.2\times10^{-2},
-1.6\times10^{-2}, -2.0\times10^{-2}$, respectively.

In Fig.\ref{figAnovanish}, there exists parameter regions of
$\beta$ and $\lambda$ where the predicted $n_s$-$r$ are consistent
with the Planck constraints.
 It is noted that the predicted $r$ increases as $\lambda$
increases, but declines as $\beta$ increases. Compared with the
power-law coupling discussed in previous section, the values of $n_s$-$r$ are
actually more sensitive to the dilaton-like coupling. We can see
that the negative coupling $\beta<0$ leads to a reduced
tensor-to-scalar ratio, so that the $\phi^2$ model with $\beta<0$
can be driven to the $\sigma$ and $2\sigma$ contour, which is
consistent with the observations very well.

\subsection{Constant coupling coefficient}

For $\lambda=0$ in the previous  subsection, the model reduces to
a constant coupling case in which  \ba \label{constant}
f_1&=&f_0,~~~V=V_0\phi^n. \ea One gets \ba f=\frac{1}{3}\beta
n^2\phi^{n-2},~~~f_{\phi}=\frac{1}{3}\beta n^2\phi^{n-3}(n-2). \ea
The Hubble flow parameter $\epsilon$ is same with the previous
model, but \ba \eta=-\frac{1}{6}\beta n^3\phi^{n-4}(n-2). \ea The
spectral index $n_s$ can be written
as \ba
n_s-1&=& -n(n+2)\phi^{-2}+\frac{2\beta n^2\phi^{n-2}}{1+\frac{4}{3}\beta\phi^n},\\
\nn \ea which involves the model parameter $\beta$ in the slow-roll
approximation.

\begin{figure}[htbp]
  \centering
  \includegraphics[scale=2,width=0.7\textwidth]{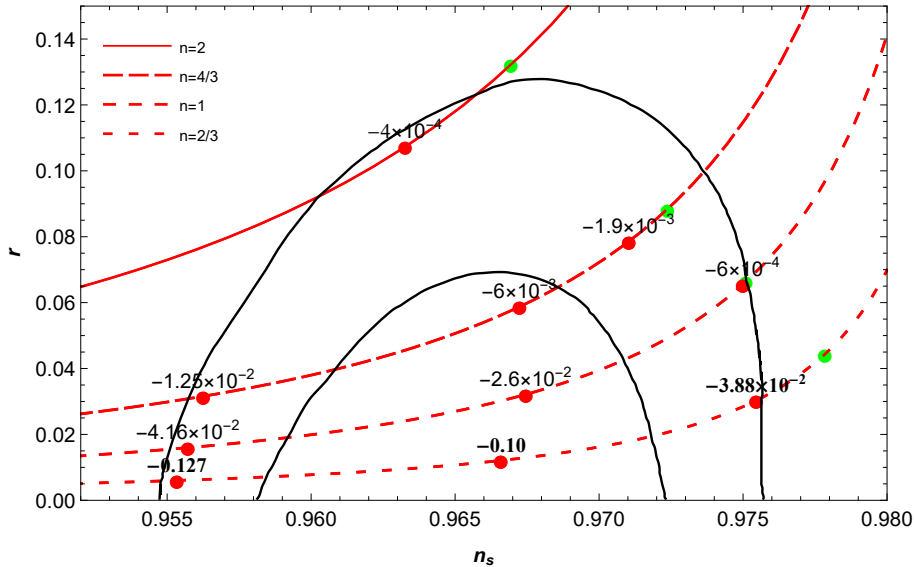}
  \caption{ The $n_s$-$r$ points predicted by the model (\ref{constant})
  with $n=2,4/3,1,2/3$ are
  confronted with Planck observation.
Here we choose $N=60$.
  The red dots corresponds to the different non-zero values of $\beta$,
  and the green dots corresponds to  the case with $\beta=0$,
  i.e.$\phi^2$ model without the $XR^{(3)}$ coupling.}\label{figAvanish}
\end{figure}

For simplicity, we choose $N=60$. In Fig.\ref{figAvanish}, we plot
 the $n_s$-$r$ predicted by the power law potentials
$\phi^n$ with $n=2, 4/3, 1, 2/3$. We can see that the
corresponding $n_s$-$r$ values can be driven to the $\sigma$ and
$2\sigma$ contours of Planck data in a suitable parameter range of
$\beta$, which is consistent with the observations.
However, contrary to the case in Fig.\ref{figbeta}, the $\phi^n$
potential with $n=4/3,1,2/3$ seems more favored than the potential
with $n=2$.


\section{Conclusion}

In this work, we have studied the slow-roll inflation with a
non-minimally derivative coupling $XR^{(3)}$.  We
work in the frame in which the graviton behaves like in standard
one, i.e. the propagating speed of gravitational waves equals to the speed of light, and analytically calculate the corrections of $XR^{(3)}$ on
the power spectra of primordial perturbations. We plot the
$n_s$-$r$ diagram  for a few inflation models with the
power-law coupling and the dilaton-like coupling, and find that
 the appearance of the $XR^{(3)}$ term can drives the
$\phi^2$ inflation models to the best-fit region of the $n_s$-$r$
diagram.
\\

\textbf{Acknowledgments}

YLH would like to thank Yong Cai and Gen Ye for helpful discussion. This work
is supported by NSFC, Nos.11575188, 11690021, and also supported by
the Strategic Priority Research Program of CAS, No.XDB23010100.

\appendix

\hypertarget{App.A}{\section{the disformal transformation of covariant action}}
\label{codisformal}

In this Appendix, we show that the covariant theory
actually belongs to the beyond Horndeski theory and is preserved
 under certain disformal
transformation.

Employing (\ref{LXR}), the covariant
action is
\ba\label{covs01} S&=&\int dx^4\sqrt{-g}\mathop{\Sigma}_{i} L_i,\\
L_2&=&G_2(\phi, X),\label{L2}\\
L_3&=&G_3(\phi, X)\Box\phi,\\
L_4&=&G_4(\phi,X)R-2G_{4X}(\phi, X)\lf(\Box\phi^2-\phi_{\mu\nu}\phi^{\mu\nu}\rt), \label{A4}\\
{}^{BH}{L}_{4} &=&  F_4(\phi,\, X) \Big[X \left( \Box\phi^2-\phi_{\mu \nu} \phi^{\mu \nu} \right)-2 \left( \Box \phi\, \phi_{\mu}\phi^{\mu\nu}\phi_{\nu} - \phi_{\mu}\phi^{\mu\nu}\phi_{\nu\rho}\phi^{\rho}\right)
\Big],\label{L5}
\ea
with
\ba
G_2&=&-\frac{X}{2}-V-\frac{f_{1\phi\phi}}{2}X^2,\\
G_3&=&-\frac{3}{2}f_{1\phi}X,\\
G_4&=&\frac{1}{2}(1+f_1X),\\
F_4&=&\frac{f_1}{2X}.
\ea
Here, we set $M_p=1$. Action (\ref{covs01}) actually belongs to a subclass of the beyond Horndeski theory \cite{Gleyzes:2014qga, Gleyzes:2014dya}.

The disformal transformation of the metric (\ref{transformation})
in covariant form reads \be \tilde
g_{\mu\nu}={C(\phi)}\lf[g_{\mu\nu}-\lf(1-\left(C(\phi)\right)^2\rt)\frac{\phi_{\mu}\phi_{\nu}}{X}\rt]
\label{discov}, \ee and the corresponding inverse
transformation is \be \tilde
g^{\mu\nu}=\frac{1}{C(\phi)}\lf[g^{\mu\nu}-\lf(1-\left(C(\phi)\right)^{-2}\rt)\frac{\phi^{\mu}\phi^{\nu}}{X}\rt],
\ee where $C(\phi)=\sqrt{1-f(\phi)},
f(\phi)=f_1\dot\phi^2\left(t(\phi)\right)$, and $g_{\mu\nu}$ is
identified as the metric in the original frame (\ref{covs01}),
while $\tilde g_{\mu\nu}$ is the metric in the new frame.

According to this transformation,
  we have \ba\label{covs02} \tilde S&=&\int d\tilde t
dx^3\sqrt{-\tilde g}\mathop{\Sigma}_{i} \tilde L_i, \ea with the
redefined coefficients \ba
\tilde G_2&=&-\frac{\tilde X}{2}-C^{-3}V+\frac{3}{4}\alpha_C^2\tilde X-\frac{1}{2}\tilde X^2(f_1C)_{\phi\phi}\\
&&+\tilde X\lf[(1+\alpha_f)\lf(\frac{\partial\alpha_C}{\partial\phi}-2\alpha_C^2\rt)-\frac{1}{2}\frac{\partial\alpha_C}{\partial\phi}\rt]\ln \tilde X,\\
\tilde G_3&=&\alpha_C(\frac{1}{2}+\alpha_f)\ln\tilde X+\alpha_C(1+2\alpha_f)-\frac{3}{2}(f_1C)_{\phi}\tilde X,\\
\tilde G_4&=&\frac{1}{2}C^{-2}+\frac{f_1}{2}C\tilde X,\\
\tilde F_4&=&\frac{1}{2\tilde X^2}(f_1C\tilde X-\alpha_f).
\ea
The definitions of $\alpha_f$ and $\alpha_C$ are given in (\ref{af}) and (\ref{alphac}), respectively.

Apparently, this new action (\ref{covs02}) maintains the same structure as (\ref{covs01}), only up to a redefinition of the coefficients. Therefore, the disformal transformation (\ref{discov}) conserves the structure of (\ref{covs01}) (see \cite{Gleyzes:2014qga} for a discussion in the unitary gauge). Action (\ref{covs01}) and action (\ref{covs02}) both belong to a subset of the beyond Horndeski theory, and suffer from the restricted conditions
\ba
G_5(\phi, \tilde X)=0,~~~F_5(\phi, \tilde X)=0.
\ea

We can rewrite the covariant action as \ba \label{tilde
s}
{\tilde S}&=&\int d\tilde t dx^3\sqrt{-\tilde g}\frac{1}{2}\Big[\tilde R+g(\phi, \tilde X)\tilde R^{(3)}-\alpha_C\lf(2+\ln\tilde X\rt)\tilde \Box\phi\nn\\
&&-\frac{\partial\alpha_C}{\partial\phi}\tilde X\ln \tilde X+\frac{3}{2}\alpha_C^2\tilde X-\tilde X-2C^{-3}V\Big],
\ea
with
\ba \label{tildeR3}
g\tilde R^{(3)}&=&\lf(\alpha_f+f_1C\tilde X\rt)\tilde R+\frac{\alpha_f+f_1C\tilde X}{\tilde X}\lf(\tilde\phi_{\mu\nu}\tilde\phi^{\mu\nu}-\Box\tilde\phi^2\rt)\nn\\
&&+2\frac{\alpha_f-f_1C\tilde X}{\tilde X^2}\lf(\Box\tilde\phi\tilde\phi^{\mu}\tilde\phi_{\mu\nu}\tilde\phi^{\nu}-\tilde\phi^{\mu}\tilde\phi_{\mu\nu}\tilde\phi^{\nu\sigma}\tilde\phi_{\sigma}\rt)\nn\\
&&+2\lf(-2\alpha_CC^{-2}\frac{1}{\tilde X}+(f_{1}C)_{\phi}\rt)\lf(\tilde \phi^{\mu}\tilde \phi_{\mu\nu}\tilde \phi^{\nu}-\tilde X\Box\tilde\phi\rt),\\
g&=&\alpha_f+f_1C\tilde X, \ea where
$\tilde\phi_{\mu\nu}=\tilde\nabla_{\mu}\tilde\nabla_{\nu}\phi$,
$\Box\tilde\phi=\tilde g^{\mu\nu}\tilde\phi_{\mu\nu}$. In unitary
gauge,  up to the second-order EFT operators, (\ref{tilde
s}) is mapped to (\ref{actioneft2}). By introducing the
time-dependent parameter
$$\alpha=-\frac{C_{\phi}}{C}\dot\phi,$$
which is equivalent to (\ref{a}), and variation with respect to the metric and $\phi$, we obtain the background equations (\ref{back01})-(\ref{phimotion}).

\end{document}